\documentclass[%
 reprint,
 amsmath,amssymb,
 aps,
prb,
]{revtex4-1}
\usepackage{subfigure}
\usepackage{graphicx}
\usepackage{dcolumn}
\usepackage{bm}
\usepackage{color} 

\begin{document}

\preprint{APS/123-QED}

\title{Vibrational effects in charge transport through a molecular double quantum dot}

\author{Jakub K. Sowa}
\author{Jan A. Mol}
\author{G. Andrew D. Briggs}
\affiliation{%
 Department of Materials, University of Oxford, Parks Road, Oxford OX1 3PH, United Kingdom
}%

\author{Erik M. Gauger}
\affiliation{SUPA, Institute of Photonics and Quantum Sciences,\\
Heriot-Watt University, EH14 4AS, United Kingdom}

\date{\today}

\begin{abstract}
Recent progress in the field of molecular electronics has revealed the fundamental importance of the coupling between the electronic degrees of freedom and specific vibrational modes. Considering the examples of a molecular dimer and a carbon nanotube double quantum dot, we here theoretically investigate transport through a two-site system that is strongly coupled to a single vibrational mode. Using a quantum master equation approach, we demonstrate that, depending on the relative positions of the two dots, electron-phonon interactions can lead to negative differential conductance and suppression of the current through the system. We also discuss the experimental relevance of the presented results and possible implementations of the studied system.
\end{abstract}

\maketitle


\section{\label{sec:level1}Introduction}
Molecular electronics has long promised reduced energy consumption, increased capability, as well as cheaper manufacturability of electronic circuits. However, the field is only now entering a new phase of research based on single-molecule devices\cite{MolE,perrin2015}. There now exist several methods which enable efficient fabrication of single-molecule junctions. Break junctions\cite{liang2002,bruot2012} and graphene nanogaps\cite{Mol1,island2014} allow single molecules to bridge the gap between a source and a drain electrodes, whilst alternative approaches rely on various scanning probe techniques\cite{guedon2012,swart2011}. This is not only paving the way towards practical molecular electronics, but also allows for experimental investigation of charge transport through complex molecular structures.

The effects of the electron-phonon coupling on the charge transport properties of single-molecule devices have been observed experimentally in a variety of molecular systems \cite{perrin2011charge,bohler2007,de2008vibrational,osorio2010conductance,Wu2004,thijssen2006} and carbon nanotube (CNT) quantum dots\cite{CNT1,lassagne2009coupling,steele2009strong,leroy2004electrical,ares2016resonant}. These differ significantly from typical solid-state quantum dots in several respects. The electronic degrees of freedom are often strongly coupled to specific vibrational modes (rather than entire phonon baths). They thus bear some resemblance to quantum-shuttles\cite{flindt2004, fedorets2004,novotny2004,novotny2003} and other mesoscopic systems\cite{okazaki2016gate,shevchenko2015delayed}. The specific impact of the electron-phonon coupling on charge transport can vary greatly from one nanoscopic system to another, depending on the number of vibrational modes and electronic states involved in the transport as well as the strength of interactions between them.  Coupling to a single vibrational mode typically results in equally spaced conductance peaks as consecutive vibrational levels enter the bias window.\cite{galperin2007molecular} Additional less trivial effects have also been demonstrated. These include: negative differential conductance (NDC), rectification, local cooling, as well as large asymmetries in the conductance maps\cite{Wu2004,hartle2009,romano2010heating,hartle2011r,galperincooling,zazunov2006,leijnse2008kinetic,koch2005effects,Shen2007,lu2015effects}. Strong electron-phonon coupling has also been experimentally shown to suppress current through single-molecule junctions\cite{FC2,FC3} and CNT quantum dots\cite{CNT1} (so-called Franck-Condon (FC) blockade\cite{Koch1}).\\ 
\begin{figure}[h]
\subfigure[\label{figure1a}\ Carbon nanotube double quantum dot]{
\includegraphics[scale=0.35]{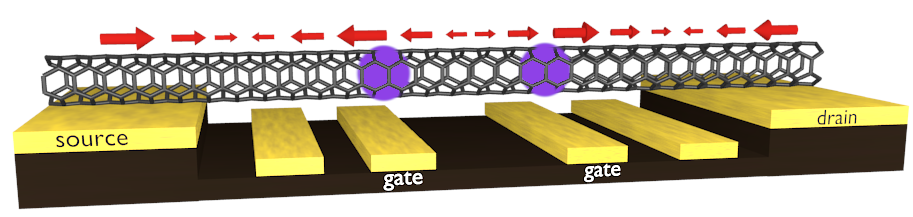}}
\subfigure[\label{figure1b}\ Molecular dimer-based single-molecule junction]{
\includegraphics[scale=0.35]{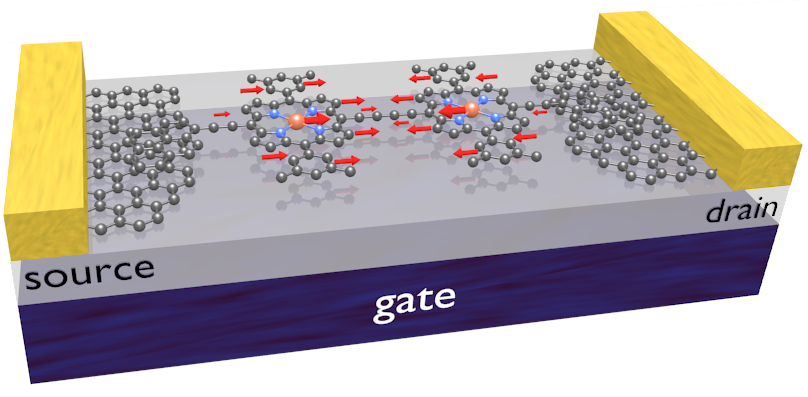}}
\caption{\label{figure1}Schematic illustrations of molecular double quantum dot systems discussed. Both sites are coupled to the same vibrational mode as well as to respective electrodes. Exemplary phonon modes are schematically depicted above. (a) shows a carbon nanotube with a pair of quantum dots (in purple) formed by applying a gate potential (as labelled). (b) depicts a single molecule junction based on a zinc-porphyrin dimer bridging a pair of graphene electrodes. Each of the porphyrin sites acts as a quantum dot.}
\end{figure}\\
Long recognised as important \cite{phonon}, the role of electron-phonon coupling in these systems has been investigated theoretically in the past using different methodologies. They are primarily based on rate equation\cite{boese2001,koch2006,koch2005effects,lu2011laserlike}, quantum master equation\cite{hubener2009master,schaller2013,hartle2011r,strasberg2016nonequilibrium,utami2004charge,muller2016dephasing,muller2016synthesising} or non-equilibrium Green's functions\cite{galperin2006,hartle2008,hartle2009,white2012inelastic,stadler2014ground,stadler2015control} methods, although other approaches have also been suggested\cite{cizek2004theory,zimbovskaya2009vibration}. Even though quantum master equation methods are typically limited to a perturbative treatment of the lead-molecule (lead-quantum dot) interactions, they provide a powerful and yet intuitive technique for investigating vibrationally-coupled quantum transport. They can be used to account for the interactions between electronic degrees of freedom and single vibrational modes, or whole phonon baths\cite{stace2005population,stace2013dynamical} (also with nontrivial spectral structure), or both of those simultaneously\cite{fruchtman2015quantum}.

Here, we study transport properties of a double quantum dot (DQD) coupled to a single vibrational (phonon) mode, depicted schematically in Fig. \ref{figure1}. Possible experimental implementations of such a system include carbon nanotube DQDs which can be nowadays almost routinely fabricated \cite{laird,PhysRevB.93.235428,biercuk2005gate} (Fig. \ref{figure1a}), and certain single-molecule junctions (Fig. \ref{figure1b}). Carbon nanotubes can be either metallic or semiconducting, depending on their chirality. In semiconducting carbon nanotubes it has been experimentally demonstrated that potential along the nanotube channel can be manipulated using electrostatic gates to define single or double quantum dot systems \cite{jorgensen2006single,mason2004local,jung2013ultraclean}. The vibrational frequencies as well as the strengths of the electron--phonon coupling usually depend on the microscopic details of the carbon nanotube, such as its chirality and radius\cite{popov2006radius,mariani2009electron}.

DQDs in single-molecule junctions can be realised by using a two-site molecular system provided the location of the charge carrier can always be well approximated as being on one of the sites (or not on the DQD at all). Several examples of such molecules have recently been investigated\cite{perrin2016gate,perrin2014large}. While the role of electron-phonon coupling in such systems remains largely unexplored experimentally, it has attracted some attention from a theoretical perspective\cite{brandes2003,walter2013,agarwalla2015,tagani2012phonon,krause2015thermodynamics}. A similar model to the one considered here has previously been studied by Santamore \textit{et al.} in Ref.~\onlinecite{santamore2013}. The authors assumed completely filled (empty) density of states on the source (drain) electrode, thermalised vibrational mode, as well as ignored the vibrational effects in lead-dot couplings. Under these assumptions, it was demonstrated that the transport through this system can be greatly enhanced if the energy difference between the sites is on resonance with the frequency of the vibrational mode.

In this work we derive a quantum master equation where we treat the quantum dot-leads coupling perturbatively, but describe the electron-phonon interactions exactly. In contrast to the work of Santamore \textit{et al.}, our theoretical treatment allows us to study the system under a finite bias as well as correctly account for the (non-equilibrium) behaviour of the vibrational mode.  Using this approach, we demonstrate that, depending on the relative phase difference in electron-phonon coupling constants between the two sites, coupling to a single bosonic mode can lead to current suppression and negative differential conductance. Finally, we discuss possible experimental implications of the presented results. Taken to the appropriate parameter regimes, we believe our model applies to both two-site molecular systems as well as CNT double quantum dots, and will provide the theoretical groundwork for experimental studies of these systems.

\section{\label{sec:level2}Model}
\begin{figure}
\includegraphics[scale=0.35]{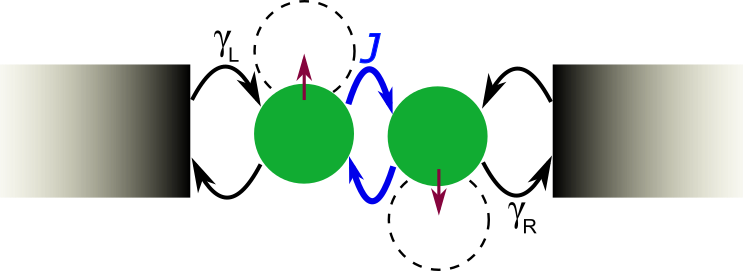}
\caption{\label{figure1c} Schematic illustration of a double quantum dot studied in this work. The two sites are coupled to each other with the strength $J$. The rates of electron hopping between the electrodes and respective sites are given by $\gamma_L$, $\gamma_R$ - see text. Note that our model  may apply to either a transverse or a longitudinal vibrational mode (as sketched in Fig. \ref{figure1}), provided the coupling strength of charge state to a single mode dominates. }
\end{figure}

In this work we consider a double quantum dot or, equivalently, a molecular dimer system in which the two sites couple to a single harmonic oscillator (which we interchangeably also refer to as a vibrational or a phonon mode). Each of the sites, hereafter denoted as left (L) and right (R), is also coupled to a respective lead constituting a fermionic reservoir, as schematically depicted in Fig. \ref{figure1c}. Henceforth, it will be assumed that $\hbar = 1$ and the electron charge $e = 1$.
\subsection{\label{sec:level21}Hamiltonian}
The total Hamiltonian for this system is given by:
\begin{equation}\label{Ham0}
H = H_0 + H_{\mathrm{e-ph}} + H_{\mathrm{coup}} + H_{\mathrm{tun}} ~,
\end{equation}
where
\begin{equation}
H_0 = H_{\mathrm{mol}} + H_{\mathrm{osc}} + H_{\mathrm{leads}} ~.
\end{equation}
$H_0$ describes the energy levels $\varepsilon_L$ and $\varepsilon_R$ of the double quantum dot ($H_{\mathrm{mol}}$), the harmonic oscillator of frequency $\omega$ ($H_{\mathrm{osc}}$), and the source and drain electrodes ($H_{\mathrm{leads}}$) as follows:
\begin{align}
&H_{\mathrm{mol}} =\varepsilon_L a^{\dagger}_L a_L + \varepsilon_R a^{\dagger}_R a_R ~,\\
&H_{\mathrm{osc}} =\omega b^\dagger b ~,\\
&H_{\mathrm{leads}} =\sum_{i,k_i}  \epsilon_{k_i} c^\dagger _{k_i} c _{k_i}
\end{align}
with $i=\{\mathrm{L},\mathrm{R}\}$.
Here, $a^{\dagger}_i$ ($a_{k_i}$) and $c^{\dagger}_i$ ($c_{k_i}$) are the creation (annihilation) operators for an electron in state $ \lvert i\rangle$ on the double quantum dot and in state $ \lvert k_i\rangle$ on the left or right lead, respectively, while $b^{\dagger}$ ($b$) is the raising (lowering) operator of the vibrational mode. The vibrational basis consists of states $\lvert v_j\rangle$ which are eigenstates of the $H_{\mathrm{osc}}$ Hamiltonian with eigenvalues $\omega\:v_j$. Here, $v_j$ is the vibrational quantum number which can take non-negative integer values.

The two dots are coupled to each other with strength $J$ as described by $H_{\mathrm{coup}}$, and to the harmonic oscillator with strengths $g_L$ and $g_R$ expressed by $H_{\mathrm{e-ph}}$: 
\begin{align}
&H_{\mathrm{coup}} =J(a^{\dagger}_L a_R + a^{\dagger}_R a_L) ~,\\
&H_{\mathrm{e-ph}} = a^{\dagger}_L a_L (g_L\: b^\dagger + g_L^{*} b) + a^{\dagger}_R a_R (g_R\: b^\dagger + g_R^{*} b) ~.\label{Hfin}
\end{align}
The electron-phonon coupling constants feature a position-dependent phase\cite{mccutcheon2011coherent,nazir2009,fassioli2010quantum,doll2006limitation}. The origin of these phases will be discussed in the next section.
Taking the position in the middle of the dots as the origin of the coordinate system, these are given by $g_L =\lvert g_L\rvert e^{-i \textbf{q}\cdot\textbf{d}/2}$ and $g_R =\lvert g_R\rvert e^{i \textbf{q}\cdot\textbf{d}/2}$, where $\textbf{q}$ is the wavevector of the phonon mode and $\textbf{d}$ is the separation between the two dots. Henceforth, the difference in phases of electron-phonon coupling will be denoted as $\xi = \textbf{q}\cdot\textbf{d}$. Finally, the operator $H_{\mathrm{tun}}$ accounts for the tunnel coupling between the sites and respective leads, 
\begin{equation}
H_{\mathrm{tun}} = \sum_{i,k_i}( V_{k_i}  c^\dagger _{k_i} a_i +  V^*_{k_i}  c _{k_i} a^\dagger_i ) ~.
\end{equation}
The assumption that the leads couple to distinct sites rather than to entire (delocalized) orbitals is well justified for tunnel coupling due to an exponential distance dependence of the tunnelling efficiency.

The electron-phonon coupling term can be eliminated from the above Hamiltonian by performing the polaron (Lang-Firsov) transformation\cite{brandes,mahan2013many}:
\begin{equation}
\bar{H} = e^{S}H \:e^{-S} ~,
\end{equation}
where $S = a^{\dagger}_L a_L\; \omega^{-1}(g_L\: b^\dagger - g_L^{*} b) + a^{\dagger}_R a_R\; \omega^{-1}(g_R\: b^\dagger - g_R^{*} b)$. The transformed Hamiltonian then becomes:
\begin{align}
&H = \bar{H}_{\mathrm{mol}} + H_{\mathrm{osc}} + H_{\mathrm{leads}} + \bar{H}_{\mathrm{coup}} + \bar{H}_{\mathrm{tun}} ~;\\
&\bar{H}_{\mathrm{mol}} =\bar{\varepsilon}_L a^{\dagger}_L a_L + \bar{\varepsilon}_R a^{\dagger}_R a_R ~;\\
&\bar{H}_{\mathrm{coup}} =  J(X^{\dagger}_L X_R a^{\dagger}_L a_R + X^{\dagger}_R X_L a^{\dagger}_R a_L)\label{HT} ~;\\
&\bar{H}_{\mathrm{tun}} = \sum_{i,k_i}( V_{k_i}  X_i c^\dagger _{k_i} a_i +  V^*_{k_i} X^{\dagger}_i c _{k_i} a^\dagger_i ) ~.
\end{align}
The energy levels of the two sites are now renormalized (polaron-shifted) such that $\bar{\varepsilon}_i = \varepsilon_i - \lvert g_i\rvert^2 / \omega$ and the factors $X_i^\dagger$ ($X_i$) are the displacement operators:
\begin{equation}
X_i^\dagger = \exp\left(\dfrac{g_i}{\omega}\;b^\dagger - \dfrac{g_i^*}{\omega}\;b\right) ~,
\end{equation}
which can be evaluated explicitly in the vibrational basis\cite{barnett2002} (see also discussion in Appendix \ref{AppC}).

\subsection{\label{sec:level22}Phase difference in electron-phonon couplings}

As introduced above, we consider coupling elements for the two dots differing by a phase factor $\xi = \textbf{q}\cdot\textbf{d}$. Since most the interesting effects we present in this work arise from these phases, we here briefly sketch an intuitive derivation of their origin and remark on their experimental relevance for the two model systems considered in this work.

Generically, the Hamiltonian accounting for coupling between vibrational modes and charge carriers can be written as\cite{mahan2013many}:
\begin{equation}
H_{\mathrm{e-ph}} = \sum_{\textbf{q}} M_{\textbf{q}} \: \varrho({\textbf{q}}) \: (b_{\textbf{q}} + b_{-\textbf{q}}^\dagger)~,
\label{eq:eqephon}
\end{equation}
where $M_{\textbf{q}}$ is a coupling element characterising the nature of the interaction. In a typical solid state setting, this can be obtained from first principles (see Ref.~\onlinecite{mahan2013many} for bulk piezoelectric and deformation potential coupling, or Ref.~\onlinecite{PhysRevB.93.235428} for the case of carbon nanotubes). The dependence of the electron-phonon coupling on the microscopic details of the carbon nanotubes can be absorbed into the effective coupling constants in the Hamiltonian \eqref{eq:eqephon}\cite{mariani2009electron}, as verified by a number of experimental studies\cite{CNT1,sapmaz2006tunneling,PhysRevB.93.235428}. We therefore proceed under the assumption that all microscopic details of the CNT systems we consider here (that are relevant in the Coulomb regime) can be adequately mapped onto the Hamiltonian \eqref{Ham0}. By contrast, in a molecular setting ab-initio methods such as DFT might provide an informed choice\cite{frederiksen2007inelastic}, or one may resort to an effective coupling strength extracted from experimental data. 

Since we only need to consider additional transport electrons (charge carriers), the charge density operator in Eq.~(\ref{eq:eqephon}), $\varrho({\textbf{q}})$, is given by:
\begin{equation}
\varrho({\textbf{q}}) = \sum_i a_i^\dagger a_i \int \mathrm{d}\textbf{r}\: e^{-i\textbf{q}\cdot\textbf{r}} \Psi_i(\textbf{r})^\dagger \: \Psi_i(\textbf{r})~,
\end{equation}
where $a_i^\dagger$ ($a_i$) is the creation (annihilation) operator for a charge carrier on $i$-site. Here, we have assumed that the wavefunctions are orthogonal, $\langle \Psi_i(\textbf{r})\rvert \Psi_j(\textbf{r})\rangle = \delta_{ij}$.\\

In the case of a double quantum dot the electron-phonon interaction Hamiltonian can then be written as:
\begin{equation}\label{e-ph}
H_{\mathrm{e-ph}} =\sum_{\textbf{q}}  (g_{L,\textbf{q}} \: a_L^\dagger a_L + g_{R,\textbf{q}} \: a_R^\dagger a_R) (b_{\textbf{q}} + b_{\textbf{-q}}^\dagger)~,
\end{equation}
where the coupling elements are given by $M_{i,\textbf{q}}$ multiplied by the Fourier transform of the electron density on the site $i$:
\begin{equation}
g_{i,\textbf{q}} = M_{i,\textbf{q}}\: \mathcal{P}[\Psi_i(\textbf{r})]~.
\end{equation}
If the wavefunctions $\Psi_R$ and $\Psi_L$ are identical but centred at $\pm \textbf{d}/2$, respectively, the coupling elements become by the shift property of the Fourier transform\cite{gauger2008robust}:
\begin{align}
g_{L,\textbf{q}} &= e^{-i\textbf{q}\cdot\textbf{d}/2} M_{i,\textbf{q}}\: \mathcal{P}[\Psi(\textbf{r})]~,\\
g_{R,\textbf{q}} &= e^{i\textbf{q}\cdot\textbf{d}/2} M_{i,\textbf{q}}\: \mathcal{P}[\Psi(\textbf{r})] ~,
\end{align}
and in the case of symmetric electron-phonon coupling $\lvert g_{L,\textbf{q}}\rvert  = \lvert g_{R,\textbf{q}}\rvert = g_{\textbf{q}}$ as:
\begin{align}
g_{L,\textbf{q}} &= e^{-i\textbf{q}\cdot\textbf{d}/2}g_{\textbf{q}}~,\\
g_{R,\textbf{q}} &= e^{i\textbf{q}\cdot\textbf{d}/2}g_{\textbf{q}} ~.
\end{align}
Finally, since the summation in equation \eqref{e-ph} runs across the entire Brillouin zone, the coupling Hamiltonian can be more compactly  written as:
\begin{equation}\label{Hq}
H_{\mathrm{e-ph}} =\sum_{i,\textbf{q}}  a_i^\dagger a_i\: (g_{i,\textbf{q}}^* b_{\textbf{q}} + g_{i,\textbf{q}} b_{\textbf{q}}^\dagger) ~.
\end{equation}
Under the assumption that the electronic degrees of freedom predominantly couple to a single phonon mode, the Hamiltonian \eqref{Hq} reduces to the one in equation \eqref{Hfin}.

Since the choice of origin is arbitrary, the important quantity in the transport through the DQD is not the phase of electron-phonon coupling itself but rather the phase difference between the two sites. Consider for example a longitudinal mode propagating through a carbon nanotube, as schematically shown in Fig. \ref{figure1a}. Difference in phases is given by $\xi = \textbf{q}\cdot\textbf{d}$ which in this case reduces to:
\begin{equation}
\xi = \lvert\textbf{q}\rvert \times\lvert\textbf{d}\rvert = \dfrac{2\pi d}{\Lambda},
\end{equation}
where $\Lambda$ is the wavelength of the relevant phonon. Thus, what determines the value of $\xi$ is the separation between the two sites in relation to $\Lambda$. In the case of CNT double quantum dots it may be possible to change the value of $\xi$ by shifting the relative positions of the two dots by using different electrostatic gate electrodes. 
Changing the distance between the two sites (with respect to the phonon mode) is not possible in single-molecule junctions. There, $\xi$ is an immutable property of a given vibrational mode.

\subsection{\label{sec:level23}Quantum Master Equation}
We proceed to trace out the degrees of freedom associated with the source and drain electrodes within the Born-Markov approximation\cite{OQS}. In other words, the dot-lead coupling is treated as a second-order perturbation. This approach is justified here since the electron-phonon coupling is larger than or comparable to the $V_{k_i}$ matrix elements\cite{mccutcheon2011general,pollock2013multi}.
Let us also remark here that no approximation has been made with regards to the interdot coupling.
As shown in the Appendix \ref{AppA}, this leads to the following master equation description of the dynamics of the double quantum dot and phonon system:
\begin{widetext} 
\begin{multline}\label{QME}
\dfrac{\mathrm{d}\bar{\rho}}{\mathrm{d}t} = - i \left[\bar{H}_{\mathrm{S}},\bar{\rho}(t) \right] 
+ \sum_{i} \dfrac{\gamma_i}{2} \left(  a^\dagger_i W_i^\dagger \bar{\rho}(t)a_i X_i + a^\dagger_i X^\dagger_i \bar{\rho}(t) a_i W_i - a_i X_i a_i^\dagger W_i^\dagger \bar{\rho}(t) - \bar{\rho}(t)a_i W_i a_i^\dagger X_i^\dagger \right)\\
+ \sum_{i} \dfrac{\gamma_i}{2} \left(  a_i Y_i \bar{\rho}(t)a_i^\dagger X_i^\dagger + a_i X_i \bar{\rho}(t) a_i^\dagger Y_i^\dagger - a_i^\dagger X_i^\dagger a_i Y_i \bar{\rho}(t) - \bar{\rho}(t)a_i^\dagger Y_i^\dagger a_i X_i \right) ~,
\end{multline}
\end{widetext}
where $\bar{H}_{\mathrm{S}} = \bar{H}_{\mathrm{mol}} + H_{\mathrm{osc}}+ \bar{H}_{\mathrm{coup}}$. The first term in equation \eqref{QME} describes the coherent evolution within the double quantum dot while subsequent terms account for incoherent hopping on and off the DQD at rates $\gamma_i = 2\pi \lvert V_i\rvert^2$. The matrix elements of the $W_i$ and $Y_i$ operators in the vibrational basis are given by: $\langle v_m \rvert W_i \lvert v_p\rangle = f_i (\bar{\epsilon}_i + \omega (v_m-v_p))\langle v_m \rvert X_i \lvert v_p\rangle$ and $\langle v_m \rvert Y_i \lvert v_p\rangle = (1 - f_i (\bar{\epsilon}_i + \omega (v_m-v_p))) \langle v_m \rvert X_i \lvert v_p\rangle$. Here, $f_i(\epsilon)$ denotes the Fermi distribution function for the $i$-lead, $f_i(\epsilon) = 1/(e^{(\epsilon - \mu_i)/kT} +1)$ and the chemical potential of the leads is determined by the applied bias voltage $V_b$: $\mu_L = + V_b/2$ and $\mu_R = - V_b/2$, respectively.
We assume that only one additional electron can be found on the DQD at any given time due to strong electron-electron repulsion. This is the so-called sequential tunnelling regime. Instead of including an explicit electron-electron interaction term, this is ensured by excluding the multiply charged states from the Hilbert space for the electronic degrees of freedom.
It then spans only three (orthogonal) states $ \lvert \mathrm{L}\rangle, \lvert \mathrm{R}\rangle$ and $ \lvert \mathrm{E}\rangle$ corresponding to an electron occupying left or right dot, and the DQD being empty, respectively. The overall ME is also trace-preserving $\mathrm{Tr}[\bar{\rho}(t)] = 1$, so that the the maximum electron population of the DQD is indeed equal to 1.


\subsection{\label{sec:level24}Electric Current}
To determine the value of the stationary electric current flowing through the system one has to first solve quantum master equation \eqref{QME} in the steady-state limit:
\begin{equation}\label{SS}
\dfrac{\mathrm{d}\bar{\rho}_{\mathrm{stat}}}{\mathrm{d}t} = 0
\end{equation}
and then compute the expectation value of the current operator in the stationary state:
\begin{equation}
I = \langle I_i\rangle_{\mathrm{stat}}= \mathrm{Tr}(I_i \bar{\rho}_{\mathrm{stat}} ).
\end{equation}
Here $I_i$ is the operator for the current flowing through the $i$-electrode. Due to current conservation, the current flowing through the left and right electrode is identical so that $I=\langle I_L\rangle_{\mathrm{stat}} = \langle I_R\rangle_{\mathrm{stat}}$. An explicit form of the current operators is given in the Appendix \ref{AppB}.

\section{\label{sec:level3}Results and Discussion}
The time-evolution of the studied system has to be described in a truncated Hilbert space. We have included $n=100$ vibrational states, which yields numerical convergence. 
Equation \eqref{SS}, given in a Hilbert space of dimension $3n\times 3n$, is solved using Krylov-subspace techniques in a way analogous to the one described by Flindt \textit{et al.} \cite{flindt2004}. The method amounts to solving equation \eqref{SS} in a subspace with reduced dimension $9n^2 \times j$ ($j$ is typically around 40) by means of Arnoldi iteration, avoiding an explicit evaluation of the Liouvillian\cite{golub1996,greenbaum1997,nation2015steady}. Generally, preconditioning is required to produce correct results. The outcomes of these calculations were tested against increasing the dimension of the Krylov subspace ($j$) and by comparing them to the results obtained using a direct method in smaller vibrational Hilbert spaces.

We distinguish two different parameter regimes depending on the relative magnitudes of dot-lead ($\gamma_L, \gamma_R$) and inter-dot couplings ($J$). First, a strong inter-dot coupling regime where the coupling between the sites is much stronger than between the dots and the leads ($J\gg\gamma_L, \gamma_R$) and second, a weak inter-dot coupling regime where the opposite is true ($\gamma_L, \gamma_R > J$). We will describe the transport properties in both these regimes in the following subsections.
\subsection{\label{sec:level31}Strong inter-dot coupling regime}
\begin{figure}[h!]
\subfigure[\label{figure2a}\ Current-voltage characteristics]{
\includegraphics[scale=0.224]{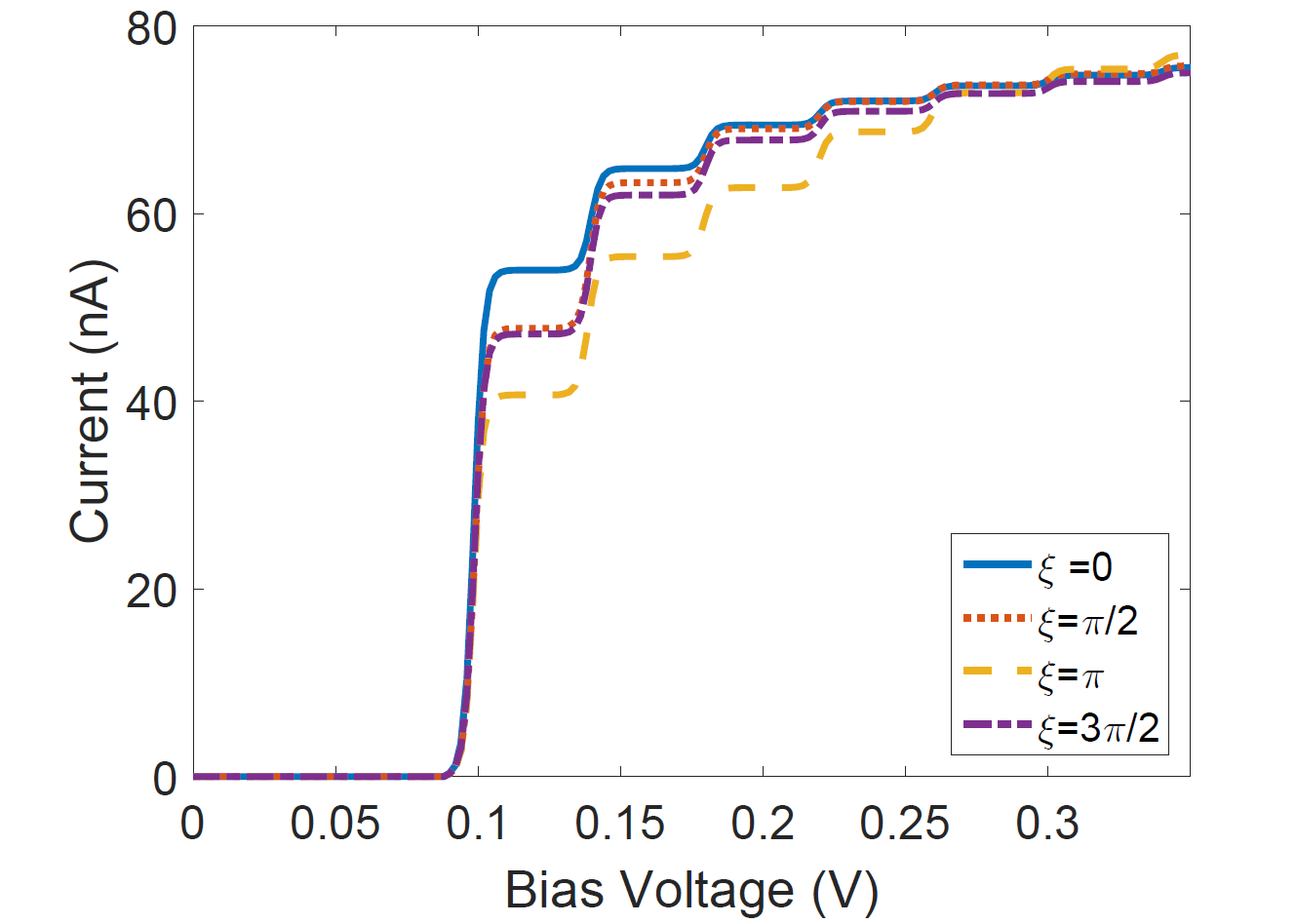}}
\subfigure[\label{figure2b}\ Populations of the sites as a function of bias voltage]{
\includegraphics[scale=0.224]{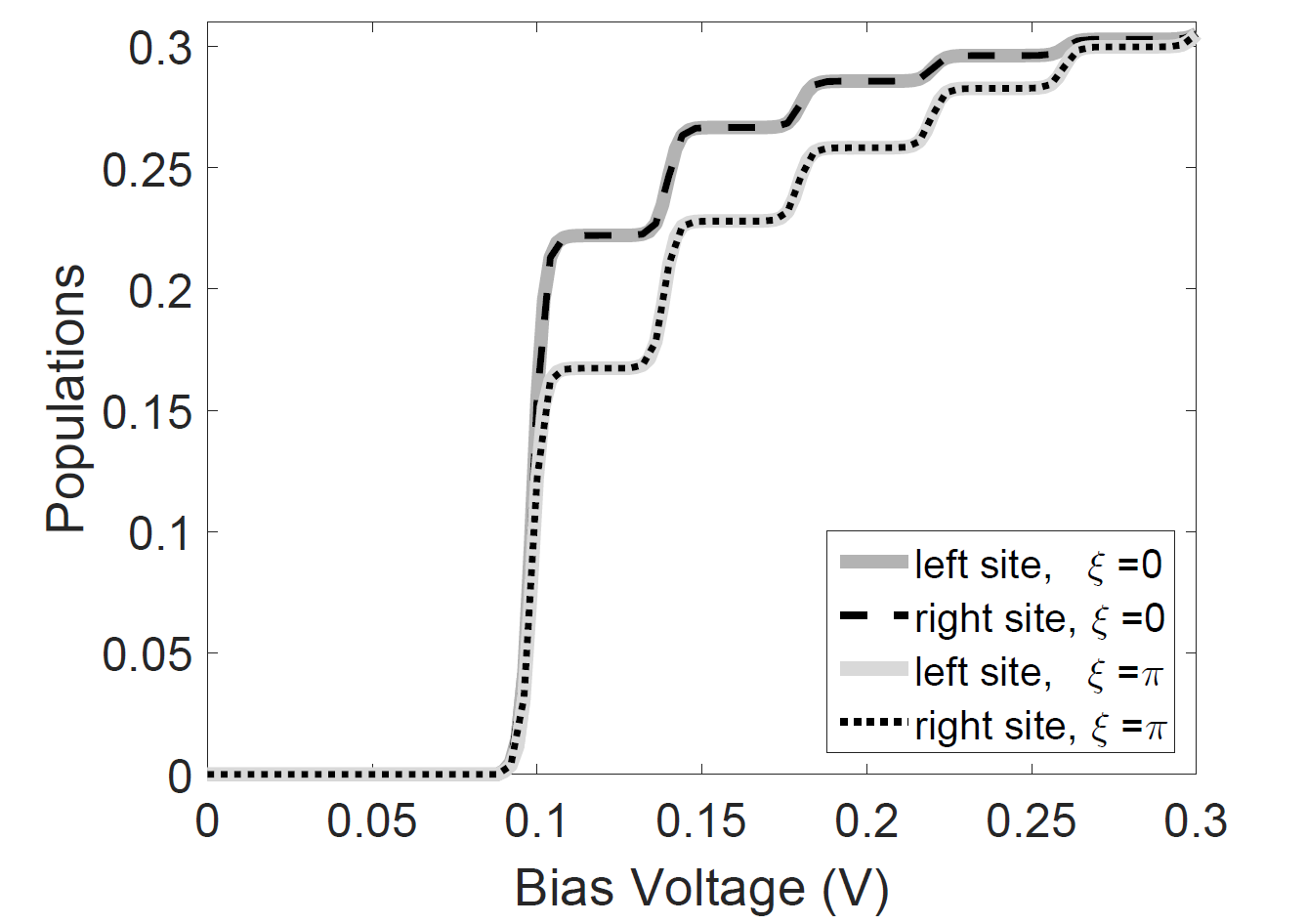}}
\caption{\label{figure2} Current-voltage characteristics  and voltage-dependent populations of the sites in a strongly coupled two-site molecular system. The calculation was performed for a symmetric system: $\bar{\varepsilon}_L = \bar{\varepsilon}_R = 0.05$ eV. The energy of the vibrational mode was taken to be $\omega = 0.02$ eV, the electron-phonon coupling constants  $\lvert g_L\rvert = \lvert g_R\rvert =0.018$ eV, the coupling between the sites $J=0.3$ eV\cite{kocherzhenko2009mechanism} and between the sites and the leads as $\gamma_L = \gamma_R=0.001$ eV. The temperature was assumed to be 10 K leading to appreciable thermal broadening.}
\end{figure}

In this section we study a double quantum dot system that is weakly coupled to the source and drain electrodes (as compared to inter-dot coupling $J$). Figure \ref{figure2a} shows the $I-V$ characteristics for such a system. For simplicity we consider a completely symmetric case such that $\bar{\varepsilon}_L = \bar{\varepsilon}_R$, $\lvert g_L \rvert = \lvert g_R \rvert$, and $\gamma_L=\gamma_R$. As expected, the system with strong inter-dot coupling behaves similarly to a single quantum dot (single-site molecular junction) coupled to a vibrational mode. The onset in the $I-V$ trace occurs when the chemical potential of the source electrode is equal to the polaron-shifted energy of the left dot $\bar{\varepsilon}_L$. The current through the system then  increases in a stepwise manner as consecutive vibrational levels fall within the bias window. The peaks in the differential conductance are separated by $2\omega$ and the relative heights of these steps can be explained qualitatively by considering the values of the $\lvert \langle v_m\rvert X\lvert 0\rangle \rvert^2$ elements (accounting for the Franck-Condon overlap between the vibrational groundstate and the $m^{\text{th}}$ vibrational excited state)\cite{zazunov2006,Koch1,hartle2011r}.

Unlike in the case for a single-site molecule, however, the phase difference in the electron-phonon coupling constant between the sites can result in an overall current suppression, especially for low vibrational levels, as shown in Fig. \ref{figure2a}. It is important to stress that the incoherent hopping (on and off the leads) is independent of the phase of electron-phonon coupling. The reason the transport is retarded in the presence of phase difference in electron-phonon coupling between the sites can be explained as follows: when the two sites couple to the vibrational mode with the same phase (so that $\xi = 2\pi$ or a multiple thereof) the factor $X_L^\dagger X_R$ in the inter-dot coupling Hamiltonian $\bar{H}_{\mathrm{coup}}$ [see Eq.~\eqref{HT}] becomes an identity operator on the vibrational space, $X_L^\dagger X_R=\openone$. This means that coherent transitions between the sites take place without any vibrational excitations. However, if the sites couple to the vibrational mode with different phases this is no longer the case and the coherent evolution within the DQD will be accompanied by phonon emission. Higher vibrational states produced in such a way will then be less efficient in tunnelling out to the leads due to poorer Franck-Condon overlap leading to an overall suppression of the current. Interestingly, at large bias voltage, the trend can reverse such that coupling in antiphase results in a marginally higher current. There, the higher vibrational states are formed during incoherent hopping of an electron from the leads onto the DQD and the coherent transition between the sites can increase the populations of lower-lying vibrational states. However, this subtle effect may be difficult to observe in practical situations. 

As shown in Fig. \ref{figure2b}, the populations on the two sites are equal independently of the applied bias and of phase difference in the electron-phonon coupling. This should come as no surprise given that the  two sites are strongly coupled and the system is entirely symmetric. 
\begin{figure}
\includegraphics[scale=0.224]{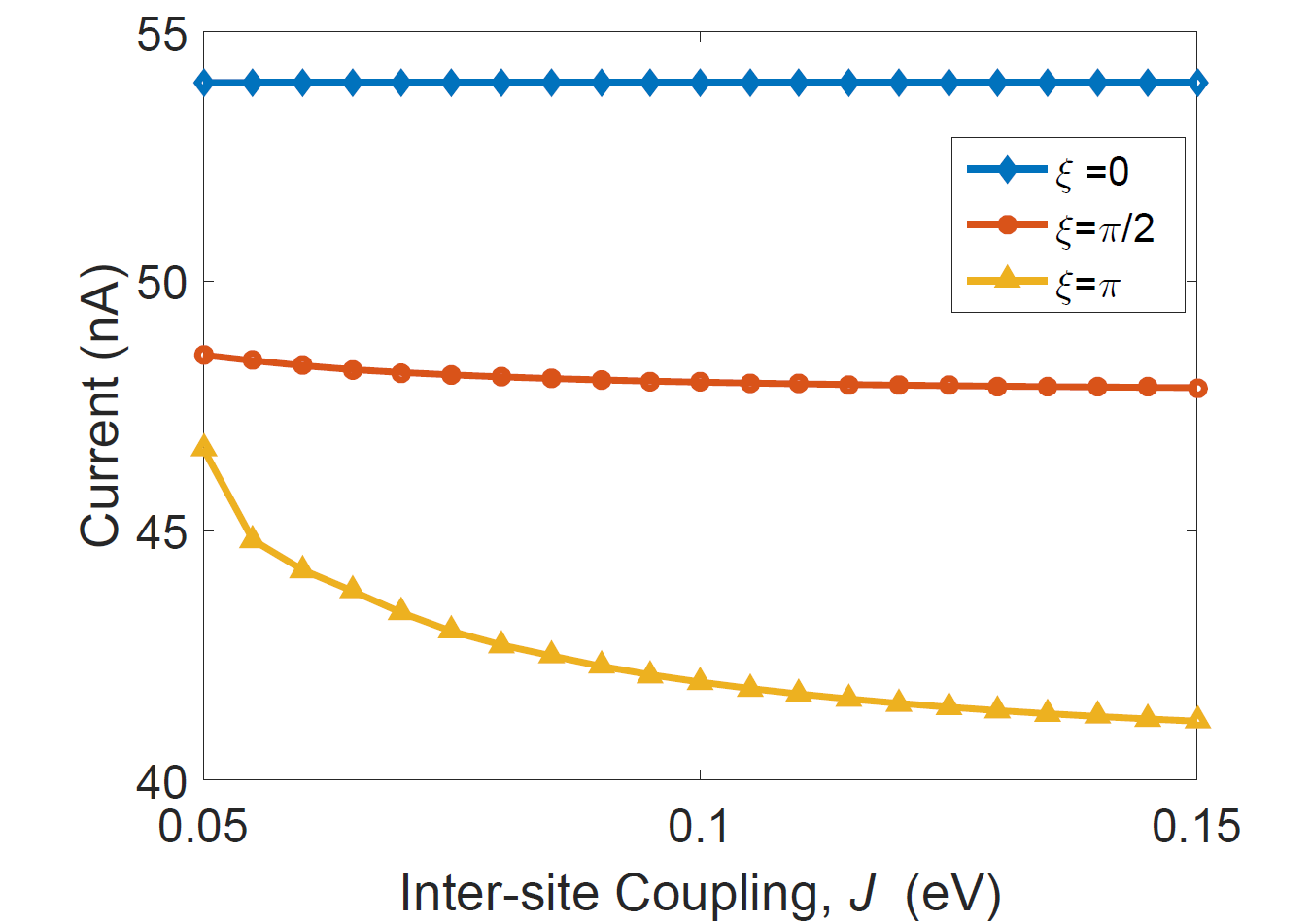}
\caption{\label{figure3} Current at bias voltage $V_b = 2\bar{\varepsilon}_L + \omega$ (first step in the current-voltage characteristics). The calculation was performed for a symmetric system: $\bar{\varepsilon}_L = \bar{\varepsilon}_R = 0.05$ eV. The energy of the vibrational mode was taken to be $\omega = 0.02$ eV, the electron-phonon coupling constants  $\lvert g_L\rvert = \lvert g_R\rvert =0.018$ eV, the coupling between the sites, and the leads is $\gamma_L = \gamma_R=0.001$ eV and the temperature was assumed to be 10 K.}
\end{figure}

Let us now consider the dependence of the current through the molecule (at the bias voltage $V_b = 2\bar{\varepsilon}_L + \omega$) on the strength of inter-site coupling $J$ in the limit $J \gg \gamma_L, \gamma_R$. In the case of $\xi=0$ the transport is virtually independent of $J$ as it is limited by incoherent hopping rates $\gamma_i$, as demonstrated in Fig. \ref{figure3}. However, as discussed above, when $\xi$ is nonzero the internal dynamics of the system becomes relevant. For $\xi=\pi/2$ and $\xi=\pi$, increasing the strength of the inter-site coupling $J$ leads to a reduction of current. This surprising result can be understood by recalling that for nonzero $\xi$ the coherent transition between the sites `produces' higher vibrational states which then lead to the observed current suppression. The stronger the coupling $J$, the greater the degree of vibrational excitation which accumulates before the electron tunnels out into the leads.

Finally, let us stress that one should expect this parameter regime (where the inter-dot coupling $J$ is stronger than the coupling to the leads) to be appropriate in the case of most single-molecule junctions\cite{kocherzhenko2009mechanism,perrin2014large}. The strength of the electron-phonon coupling in these systems is often estimated experimentally by considering the relative heights of the Franck-Condon steps\cite{CNT1,FC3}. As shown above, one has to be careful in using the same approach for two- or multiple-site systems, where the phase difference in electron-phonon coupling becomes relevant. Another surprising conclusion from the presented results is that the intramolecular dynamics can quite significantly affect the transport properties of the studied system in this parameter regime, despite the fact that the transport efficiency is limited by hopping between the leads and the molecular system.
\subsection{\label{sec:level32}Weak inter-dot coupling regime}
\begin{figure}[h!]
\subfigure[\label{figure4a}\ Current-voltage characteristics]{
\includegraphics[scale=0.224]{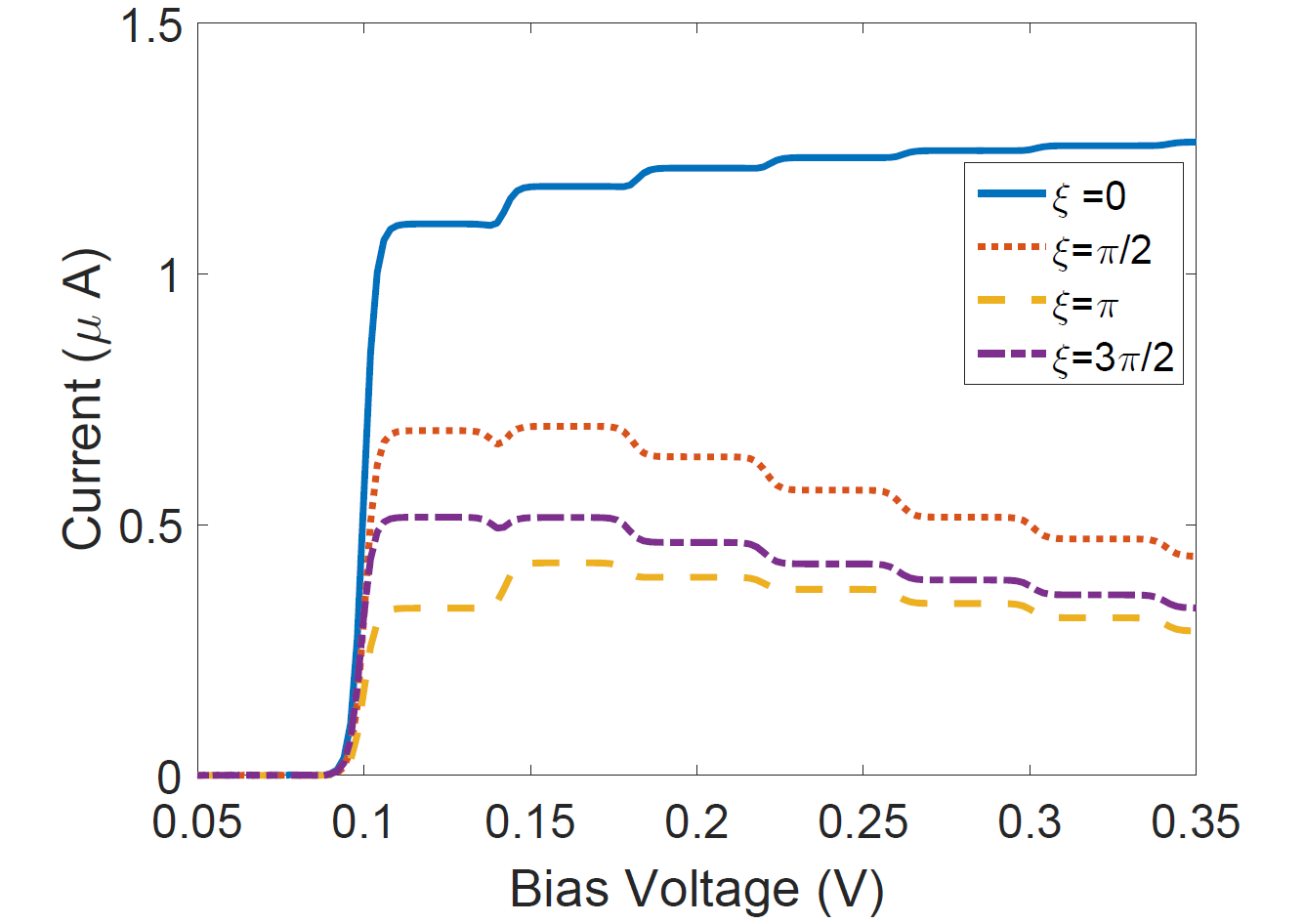}}
\subfigure[\label{figure4b}\ Populations of the sites as a function of bias voltage]{
\includegraphics[scale=0.224]{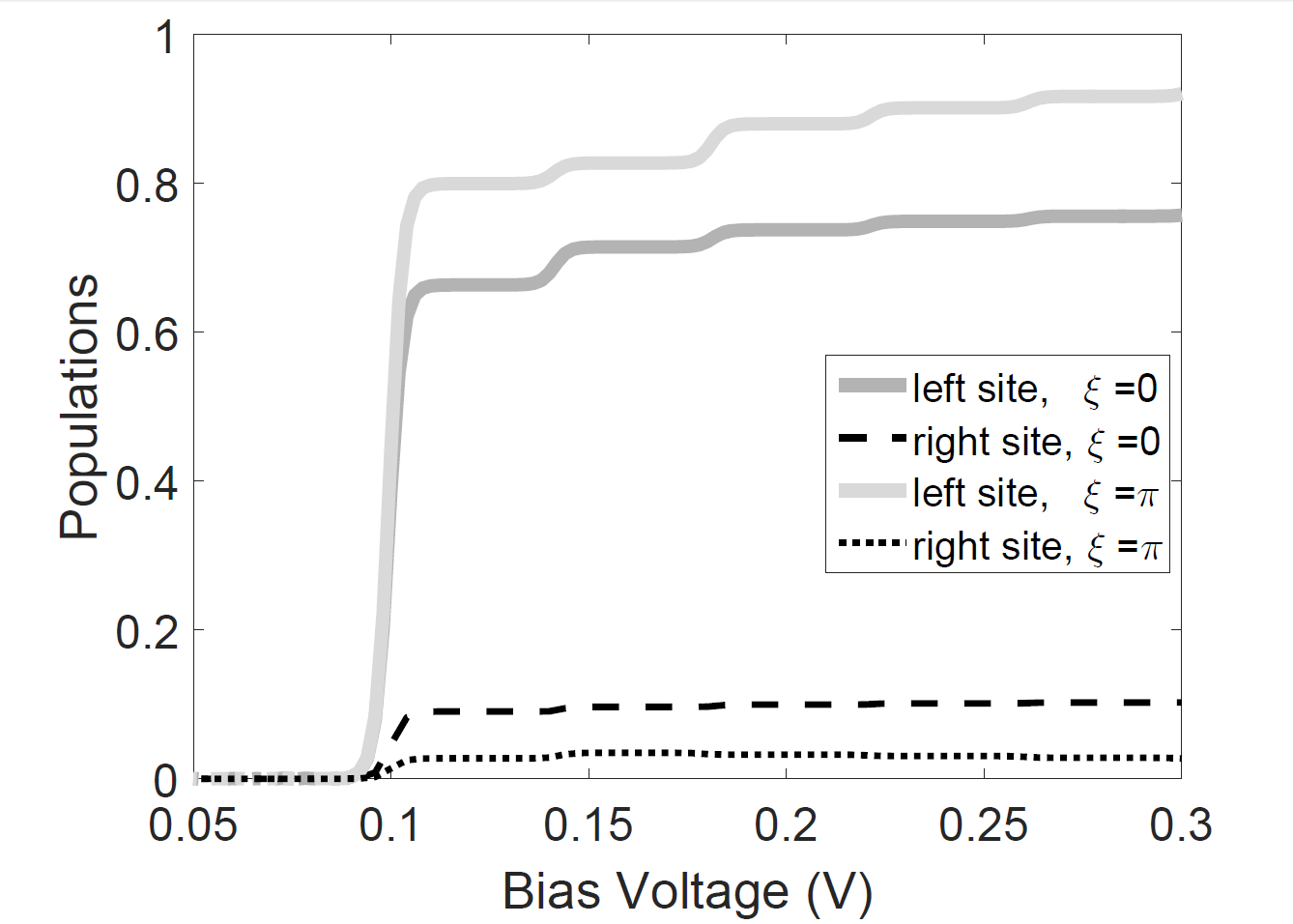}}
\caption{\label{figure4} Current-voltage characteristics  and voltage-dependent populations of the sites in a strongly coupled two-site molecular system. The model is identical to the one used in Fig. \ref{figure2} except for the coupling between the sites, $J=0.01$ eV, and between the sites and the leads: $\gamma_L = \gamma_R=0.05$ eV.}
\end{figure}

Let us now consider the reverse situation, that is, when the two dots are weakly coupled to each other but interact with the electrodes more strongly (this parameter regime can be realised in CNT double quantum dots, see for example Ref.~\onlinecite{steele2009tunable}). As we will now discuss, this case delivers very different and much richer physical phenomena. Once again, we shall first focus on a symmetric system. As before, when there is no phase difference in the vibrational coupling between the sites, one observes a stepwise increase in the current through the system, see Fig. \ref{figure4a}. This can again be explained by vibrational effects in the tunnel coupling between the sites and the leads.
Even though the bottleneck in the transport is the coherent hopping between the sites, vibrational effects in connection with incoherent hopping on (and off) the molecule will play an important role as long as the hopping rates $\gamma_i$ and inter-site coupling $J$ are comparable.

Moving on to the case of nonzero $\xi$, we can see that a difference in phase between the electron-phonon coupling elements still induces a current suppression. In addition, the phase difference leads to negative differential conductance -- a decrease in current with increasing bias voltage. In this parameter regime, it is the inter-site transition that limits the overall transport rate. Hence, what is crucial now is how the phase difference in electron-phonon couplings affects the $\lvert L\rangle \leftrightarrow \lvert R\rangle$ transition. It turns out that this transition is suppressed for nonzero $\xi$, resulting in the observed current suppression, and is even less efficient for higher vibrational states (as discussed below), giving rise to the negative differential conductance shown in Fig. \ref{figure4a}.

Since it is the coherent hopping between the sites that is the rate-limiting step in the overall charge transport, the populations of the two sites are no longer equal (Fig. \ref{figure4b}). In the case of $\xi=0$, populations of the two sites increase with the applied bias voltage. However, if the sites couple to the vibrational mode with different phases, an increase in $V_b$ is accompanied by a decrease in the population of the right site (correlating with the current through the DQD).
This demonstrates a rather different physical origin of observed effects as compared with the ones reported in section \ref{sec:level31} (\textit{c.f.} Fig. \ref{figure2b}).
\begin{figure}
\subfigure[\label{figure5a}\ Current-voltage characteristics, $\xi=\pi$]{
\includegraphics[scale=0.224]{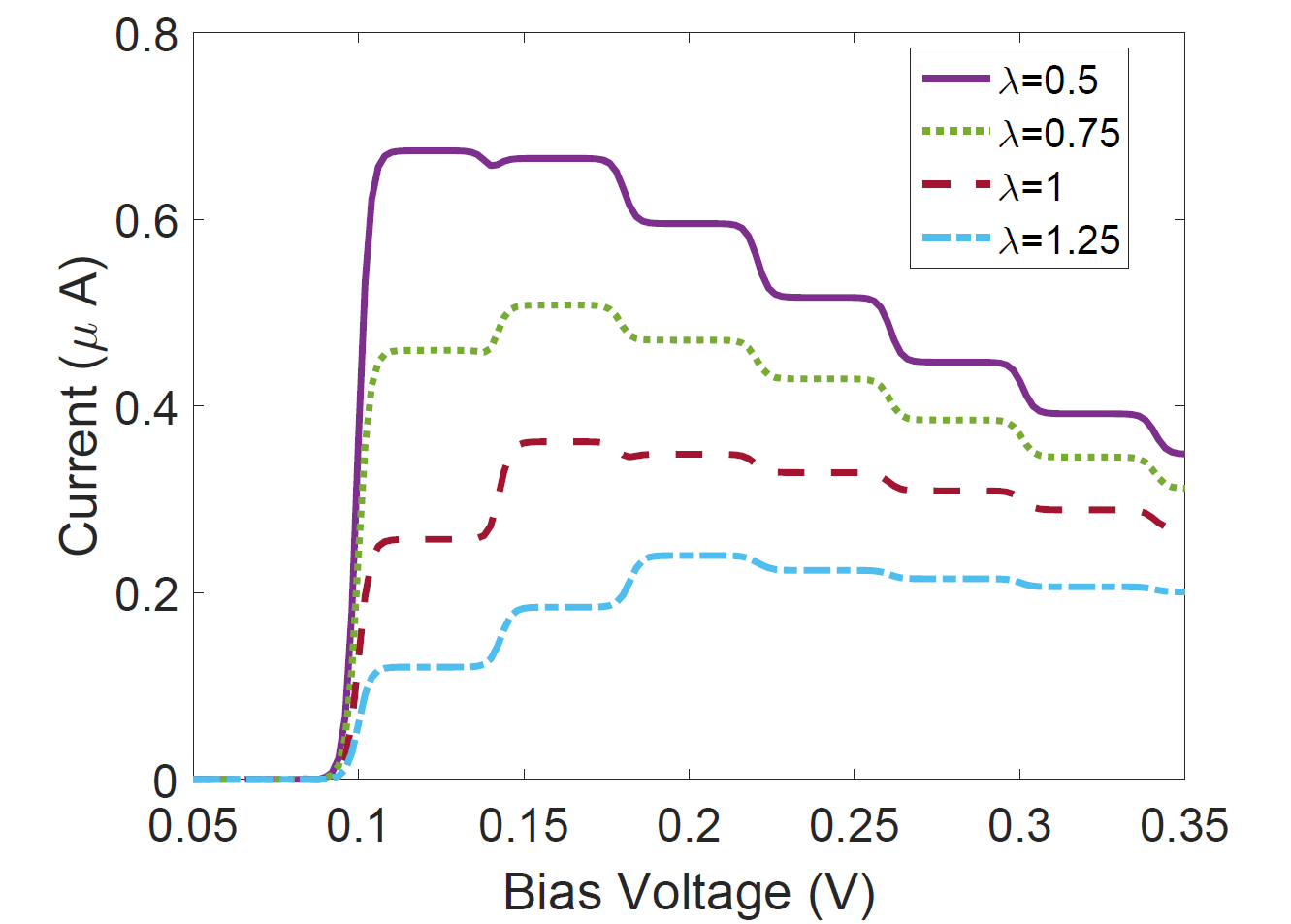}}
\subfigure[\label{figure5b}\ Current at bias voltage $V_b = 2\bar{\varepsilon}_L + \omega$]{
\includegraphics[scale=0.224]{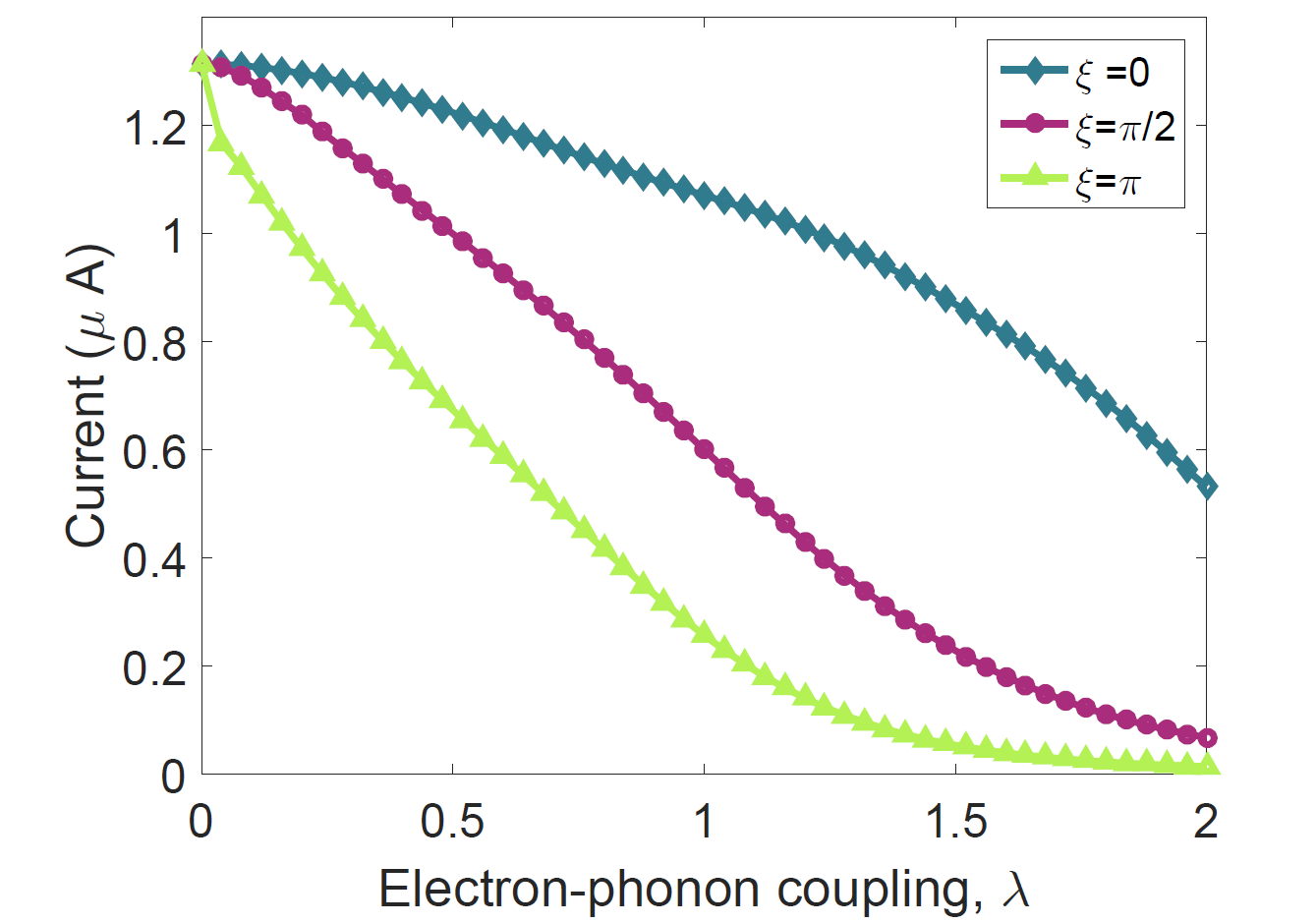}}
\caption{\label{figure5} Current-voltage characteristics (a) and current at bias voltage $V_b = 2\bar{\varepsilon}_L + \omega$ (b) for a symmetric DQD: $\bar{\varepsilon}_L = \bar{\varepsilon}_R = 0.05$. The energy of the vibrational mode was taken to be $\omega = 0.02$ eV, the electron-phonon coupling constants  $\lvert g_L\rvert = \lvert g_R\rvert =\lambda \omega$, the inter-site coupling, $J=0.01$ eV and between the sites and the leads, $\gamma_L = \gamma_R=0.05$ eV, $T=10 K$.}
\end{figure}

To understand the origin of the NDC in more detail, let us now consider transport through the two-site system for different values of electron-phonon coupling constant. Current-voltage characteristics for different values of $\lambda=\lvert g\rvert/\omega$ are shown in Fig. \ref{figure5a}. The position of the maximum value of the current shifts depending on the value of $\lambda$. This can be qualitatively related to the magnitudes of the diagonal elements of the $X_L^\dagger X_R$ operator (Fig. \ref{figureAA} in Appendix \ref{AppC}). As can be seen in Eq.~\eqref{HT} the displacement operators affect the overall efficiency of the coherent hopping between the sites. One can notice that the maximum current occurs at bias voltage 
for which the vibrational level $v$, corresponding to the largest $|  \langle v | X_R^\dagger X_L|v\rangle | ^2$ matrix element enters the bias window (compare Fig. \ref{figure5a} and Fig. \ref{figureAA}). The relative magnitudes of the steps in the current-voltage characteristics cannot be, however, as easily correlated with values of $\langle v | X_R^\dagger X_L|v\rangle$. The observed value of current is a result of a complex interplay between vibrational effects during hopping on and off the DQD and the dynamics within the system.

It is interesting to consider the value of current at the first step of the $I-V$ trace (at $V_b = 2\bar{\varepsilon}_L +\omega$) as a function of $\lambda$ and $\xi$, see Fig. \ref{figure5b}. Two overlapping effects become apparent: Firstly, a decrease in transport efficiency as the electron-phonon coupling ($\lambda$) increases - this is fundamentally an example of the Franck - Condon blockade\cite{Koch1,koch2006}. Secondly, the current decreases as the difference in phases of electron-phonon coupling $\xi$ deviates from 0 (or $2\pi$) - the effect described and explained above. 

\begin{figure}
\includegraphics[scale=0.224]{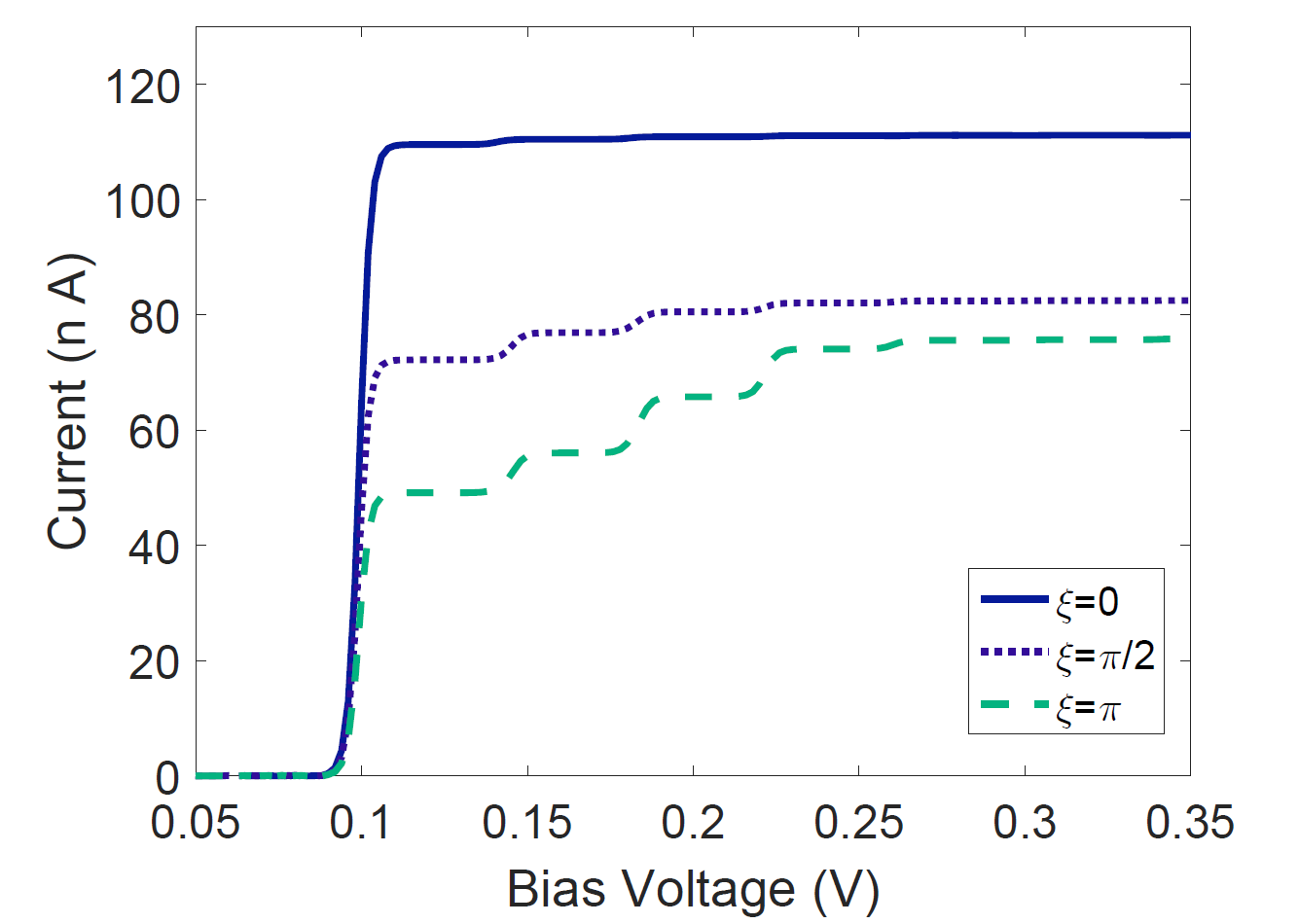}
\caption{\label{figure6} Current-voltage characteristics for a detuned DQD: $\bar{\varepsilon}_L = 0.05$ eV, $\bar{\varepsilon}_R = 0.15$ eV. The energy of the vibrational mode was taken to be $\omega = 0.02$ eV, the electron-phonon coupling constants  $\lvert g_L\rvert = \lvert g_R\rvert =0.018$ eV, the inter-site coupling, $J=0.01$ eV and between the sites and the leads, $\gamma_L = \gamma_R=0.05$ eV, $T=10 K$.}
\end{figure}

Finally, let us consider a case in which the energy levels of the two sites are detuned (for example due to an asymmetric structure of the molecule or an applied gate voltage). The current-voltage characteristics for this situation is shown in Fig. \ref{figure6}. Introducing an energy gap between the sites decreases the current through the DQD as it decreases the efficiency of $\lvert L\rangle \leftrightarrow \lvert R\rangle$ transition. Moreover, for a considerable detunning between the sites the NDC, observed in $\bar{\varepsilon}_L = \bar{\varepsilon}_R$ case, is lifted. Now, higher vibrational states (on the left site) are in fact more efficient in tunnelling from the left to right site, especially into lower vibrational states (on the right site).
Thus, increasing the bias voltage (coinciding with increasing populations of higher vibrational levels) leads to an overall increase in the current through the DQD. 

\section{\label{sec:level4}Conclusions}
In this paper, we have derived a quantum master equation that can account for a transport through a two-site molecular system (or a carbon nanotube double quantum dot) that is coupled to a single vibrational mode. 
The method used here is perturbative and relies on a Born-Markov approximation with respect to lead-dot coupling but, unlike most previous studies of similar model systems (with the notable exception of Ref. \onlinecite{krause2015thermodynamics}), treats electron-phonon interactions exactly. Using it we have demonstrated that, depending on the phase difference in electron-phonon couplings $\xi$, one can observe current suppression and/or negative differential conductance in transport through the studied system. The role of $\xi$ in vibrationally coupled charge transport remained, thus far, largely unexplored. We have also explained the physical origin of the observed effects by analysing the derived quantum master equation as well as considering the bias voltage-dependence of the on-site populations.
For simplicity, we have mostly limited our discussion to symmetric systems with $\gamma_L=\gamma_R$ and $\lvert g_L\rvert = \lvert g_R\rvert$. Considering asymmetry may well be necessary to account for future experimental observations, however, it is a trivial modification to our model presented in this work. 
We would also expect to observe similar effects to the ones described here (current suppression and NDC in relevant parameter regimes) in systems where  $\lvert g_L\rvert \ne\lvert g_R\rvert$ even in the absence of phase difference in electron-phonon coupling constants. Once again, these would be caused by vibrational excitations or suppression of the  $\lvert L\rangle \leftrightarrow \lvert R\rangle$ transition.

The phenomena described here are non-equilibrium effects and can naturally be lifted by very fast damping of the vibrational mode. Fast relaxation of the harmonic oscillator will bring the vibrational states of the system back to their ground state (at least at sufficiently low temperature) diminishing the current suppression and NDC effects described in this work. Vibrational relaxation times in CNT quantum dots and molecular junctions can be, however, quite long, as it has been observed experimentally\cite{leroy2004electrical,FC2}. One also has to bear in mind that in real systems the electronic degrees of freedom are typically coupled to more than one vibrational mode. We expect our analysis, however, to provide a reasonable description of the transport characteristics when coupling to one of the vibrational modes is dominant, as it is often the case\cite{FC2,CNT1,hartle2009,PhysRevB.93.235428}. 

Finally, we have also discussed experimental relevance of the reported findings. We believe that our model is capable of describing the behaviour of both two-site single-molecule junctions (for which strong inter-site coupling regime is most likely going to be applicable) and carbon nanotube double quantum dots (in which case the weak inter-site coupling regime may be more appropriate). While testing the theoretical predictions described here in an experiment may not be trivial, we believe that observing the effects we discuss should certainly be possible and would offer an exciting glimpse onto the rich tapestry of non-equilibrium effects in quantum transport.
\begin{acknowledgments}
The authors would like to thank Neill Lambert for useful comments.
J.K.S. also thanks the Clarendon Fund and EPSRC for financial support. E.M.G. acknowledges funding from the Royal Society of Edinburgh and the Scottish Government. This work was supported by the EPSRC QuEEN Programme Grant (EP/N017188/1).
\end{acknowledgments}
\appendix
\section{\label{AppA} Derivation of Quantum Master Equation}
The time-evolution of the density matrix for the entire system ($\bar{\chi}$) can be described in the polaron-transfered frame by:
\begin{equation}\label{QME0}
\dfrac{\mathrm{d}\bar{\chi}}{\mathrm{d}t} = -i[\bar{H},\bar{\chi}]~.
\end{equation}
In the interaction picture of $\bar{H}_0 = \bar{H}_{\mathrm{mol}} + H_{\mathrm{osc}} + H_{\mathrm{leads}}$, equation \eqref{QME0} becomes:
\begin{equation}\label{INT}
\dfrac{\mathrm{d}\tilde{\chi}(t)}{\mathrm{d}t} = -i[\tilde{H}_{\mathrm{coup}}(t) +\tilde{H}_{\mathrm{tun}}(t),\tilde{\chi}(t)]~,
\end{equation}
where
\begin{multline}
\tilde{H}_{\mathrm{tun}}(t) = e^{i\bar{H}_0 t} \bar{H}_{\mathrm{tun}} e^{-i\bar{H}_0 t}\\ = \sum_{i,k_i}\left( V_{k_i}  c^\dagger _{k_i} e^{i(\epsilon_{k_i}-\bar{\varepsilon}_i)t} a_i \tilde{X}_i(t) + \mathrm{H.c.}\right)
\end{multline}
and similarly $\tilde{H}_{\mathrm{coup}}(t) = e^{i\bar{H}_0 t} \bar{H}_{\mathrm{coup}} e^{-i\bar{H}_0 t}$ and $\tilde{X}_i(t) = e^{i\bar{H}_0 t} X_i e^{-i\bar{H}_0 t}$.

Integrating equation \eqref{INT} and substituting the solution into the second commutator, one obtains:
\begin{multline}\label{BM1}
\dfrac{\mathrm{d}\tilde{\chi}(t)}{\mathrm{d}t} =  -i[\tilde{H}_{\mathrm{coup}}(t),\tilde{\chi}(t)] -i[\tilde{H}_{\mathrm{tun}}(t),\tilde{\chi}(0)]\\  -\int_0^t \mathrm{d}s \: [\tilde{H}_{\mathrm{tun}}(t),[\tilde{H}_{\mathrm{tun}}(s) + \tilde{H}_{\mathrm{coup}}(s),\tilde{\chi}(s)]]~.
\end{multline}
We shall now make a series of assumptions commonly known together as Born-Markov approximation. Firstly, we will assume that the fermionic reservoirs interact weakly with the double quantum dot and that they always remain in thermal equilibrium. Then, the total density matrix can be written as: $\tilde{\chi}(\tau) = \tilde{\rho}(\tau)\otimes R_0 $ where $\rho(t)$ is a density matrix describing the double quantum dot and phonon system while $R_0$ accounts for the thermal state of the source and drain electrodes. Given that the second commutator vanishes, after tracing out the fermionic reservoirs equation \eqref{BM1} becomes:
\begin{multline}\label{BM2}
\dfrac{\mathrm{d}\tilde{\rho}}{\mathrm{d}t} = -i[\tilde{H}_{\mathrm{coup}}(t),\tilde{\rho}(t)]\\ -\int_0^t \mathrm{d}s \: \mathrm{Tr}_{\mathrm{leads}} [\tilde{H}_{\mathrm{tun}}(t),[\tilde{H}_{\mathrm{tun}}(s), \tilde{\rho}(s) \otimes R_0]]~,
\end{multline}
where in the terms linear in $\tilde{H}_{\mathrm{tun}}$ vanish when traced out within Born approximation\cite{brandes}~.\\
Expanding the commutators, the above can be written as:
\begin{multline}\label{BM3}
\dfrac{\mathrm{d}\tilde{\rho}}{\mathrm{d}t} =-i[\tilde{H}_{\mathrm{coup}}(t),\tilde{\rho}(t)]\\
-\sum_{i,k_i} \int_0^t \mathrm{d}s \: h_{k_{i}}(t-s)  \left(  \tilde{a}_i(t)\tilde{a}_i^\dagger (s) \tilde{\rho}(s)  - \tilde{a}_i^\dagger (s)\tilde{\rho}(s) \tilde{a}_i(t) \right) \\
 -\sum_{i,k_i} \int_0^t \mathrm{d}s \: \bar{h}_{k_{i}}(s-t) \left(  \tilde{a}_i^\dagger(t)\tilde{a}_i (s) \tilde{\rho}(s)  - \tilde{a}_i (s)\tilde{\rho}(s) \tilde{a}_i^\dagger(t) \right) \\
 -\sum_{i,k_i} \int_0^t \mathrm{d}s \: h_{k_{i}}(s-t) \left(  \tilde{\rho}(s)\tilde{a}_i(s)\tilde{a}_i^\dagger (t)  - \tilde{a}_i^\dagger (t)\tilde{\rho}(s) \tilde{a}_i(s) \right) \\ 
-\sum_{i,k_i} \int_0^t \mathrm{d}s \: \ \bar{h}_{k_{i}}(t-s) \left( \tilde{\rho}(s) \tilde{a}_i^\dagger(s)\tilde{a}_i (t) - \tilde{a}_i (t)\tilde{\rho}(s) \tilde{a}_i^\dagger(s) \right)~,\\
\end{multline}
where $ h_{k_{i}}(\tau) \equiv \lvert V_{k_i}\rvert^2\: f_i(\epsilon_{k_i}) \:e^{i\epsilon_{k_i}\tau}, \ \ \bar{h}_{k_{i}}(\tau) \equiv \lvert V_{k_i}\rvert^2\: [1 - f_i(\epsilon_{k_i})] \:e^{i\epsilon_{k_i}\tau}$
with the Fermi distribution given by: $f_i(\epsilon_{k_i}) = \mathrm{Tr}_{res} (R_0 c^\dagger _{k_i}c_{k_i}) = 1/ (e^{(\epsilon_{k_i} - \mu_i)/kT} - 1)$. Here, $\tilde{a}_i(\tau)$ ($\tilde{a}^\dagger_i(\tau)$) denotes a polaron-transformed annihilation (creation) operator in the interaction picture so that: $\tilde{a}_i(\tau) = a_i e^{-i\bar{\varepsilon}_i \tau} \tilde{X}_i(\tau)$.

We can now employ the Markov approximation which replaces $\tilde{\rho}(s)$ with $\tilde{\rho}(t)$ so that the evolution of the state at time $t$ depends only on the present state. Furthermore, we will replace $s$ with $t-s'$ and extend the upper limit of the integral to infinity, obtaining a Markovian equation that is local in time:
\begin{multline}\label{BM4}
\dfrac{\mathrm{d}\tilde{\rho}}{\mathrm{d}t} =-i[\tilde{H}_{\mathrm{coup}}(t),\tilde{\rho}(t)] +\sum_{i,k_i}
\int_0^\infty \mathrm{d}s' \\- h_{k_{i}}(s')  \left(  \tilde{a}_i(t)\tilde{a}_i^\dagger (t-s') \tilde{\rho}(t)  - \tilde{a}_i^\dagger (t-s')\tilde{\rho}(t) \tilde{a}_i(t) \right) \\
 - \bar{h}_{k_{i}}(-s') \left(   \tilde{a}_i^\dagger(t)\tilde{a}_i (t-s') \tilde{\rho}(t)  - \tilde{a}_i (t-s')\tilde{\rho}(t) \tilde{a}_i^\dagger(t) \right) \\
 -   h_{k_{i}}(-s') \left(   \tilde{\rho}(t)\tilde{a}_i(t-s')\tilde{a}_i^\dagger (t)  - \tilde{a}_i^\dagger (t)\tilde{\rho}(t) \tilde{a}_i(t-s') \right) \\ 
-   \bar{h}_{k_{i}}(s') \left(  \tilde{\rho}(t) \tilde{a}_i^\dagger(t-s')\tilde{a}_i (t) - \tilde{a}_i (t)\tilde{\rho}(t) \tilde{a}_i^\dagger(t-s') \right)~.\\
\end{multline}
The sum over the energy levels in \eqref{BM4} can be replaced with an integral. The integral over $s'$ can be performed by using the relation:
\begin{equation}
\int_{0}^{\infty} \mathrm{d}\tau e^{\pm i \Omega \tau} = \pi \delta(\Omega) \pm i \dfrac{\mathcal{P}}{\Omega}
\end{equation}
(where $\mathcal{P}$ denotes Cauchy's Principle Value) and ignoring the imaginary terms (which only lead to a minute renormalisation of the Hamiltonian)\cite{OQS}.
Performing integration over $\epsilon_{k_i}$ and then moving back to the Schr\"{o}dinger picture leads to equation \eqref{QME} given in section II of the paper.

\section{\label{AppB} Current operators}
The current operator for the current flowing through $L/R$-electrode is given by:
\begin{equation}
I_i = \dfrac{\mathrm{d}\;\lvert i\rangle\langle i\rvert}{\mathrm{d}t}~.
\end{equation}
Equivalently, due to trace-preserving property of equation \eqref{QME}, current can also be expressed as the rate of change of the empty population at the $i$-electrode:
\begin{equation}
I_i =\left.\dfrac{\mathrm{d}\;\lvert E\rangle\langle E\rvert}{\mathrm{d}t}\right\rvert_i~,
\end{equation}
which, using equation \eqref{QME}, is given by the superoperator:
\begin{multline}
I_i \bar{\rho}=  \dfrac{\gamma_i}{2} (  a^\dagger_i W_i^\dagger \bar{\rho}(t)a_i X_i + a^\dagger_i X^\dagger_i \bar{\rho}(t) a_i W_i \\
-  a_i Y_i \bar{\rho}(t)a_i^\dagger X_i^\dagger - a_i X_i \bar{\rho}(t) a_i^\dagger Y_i^\dagger )~.
\end{multline}

\section{\label{AppC} Displacement Operators}
Properties of the displacement operators are discussed in detail, for example, in Ref.~\onlinecite{brandes2003,barnett2002}. Here, we focus on the effect the phase difference in electron-phonon couplings $\xi$ has on the $X^\dagger_L X_R$ and $X^\dagger_R X_L$ operators. They can be written as:
\begin{equation}
X^\dagger_L X_R = \exp\left[ \left(\dfrac{g_L - g_R}{\omega}\right) b^\dagger -  \left(\dfrac{g_L^* - g_R^*}{\omega}\right) b\right]
\end{equation}
(disregarding the $\exp[\;i\:\mathrm{Im}(g_L g_R^*/\omega^2)]$ factor) and its Hermitian conjugate, accordingly. For the symmetric case studied in the paper $g_L = g e^{-i\xi/2}$ and $g_R = g e^{i\xi/2}$ so that:
\begin{equation}
X^\dagger_L X_R = \exp(\alpha\: b^\dagger - \alpha^* b)~,
\end{equation}
where $\alpha = 2i\dfrac{g}{\omega} \sin(\xi/2)$. The diagonal elements in the vibrational basis of this operator are given by:
\begin{equation}
\langle v | X_R^\dagger X_L|v\rangle  = \exp(-\lvert \alpha\rvert^2/2) \ L_v(\lvert \alpha\rvert^2)~,
\end{equation}
where $v$ is a vibrational quantum number and $L_v$ is a Laguerre polynomial of order $v$. From this it can be seen that although the magnitude of $\langle v | X_R^\dagger X_L|v\rangle$ depends non-trivially on $v$, due to the exponential term it should be maximum at $\xi=0$ and decrease towards $\xi=\pi$. Similarly, for a given $\xi$, its magnitude should also decrease with increasing $\lambda=\dfrac{g}{\omega}$. One can see that these trends correlate well with the conductance of the DQD (see Figure \ref{figure5b}).

\begin{figure}
\subfigure[\ $\lambda = 0.5$]{
\includegraphics[scale=0.115]{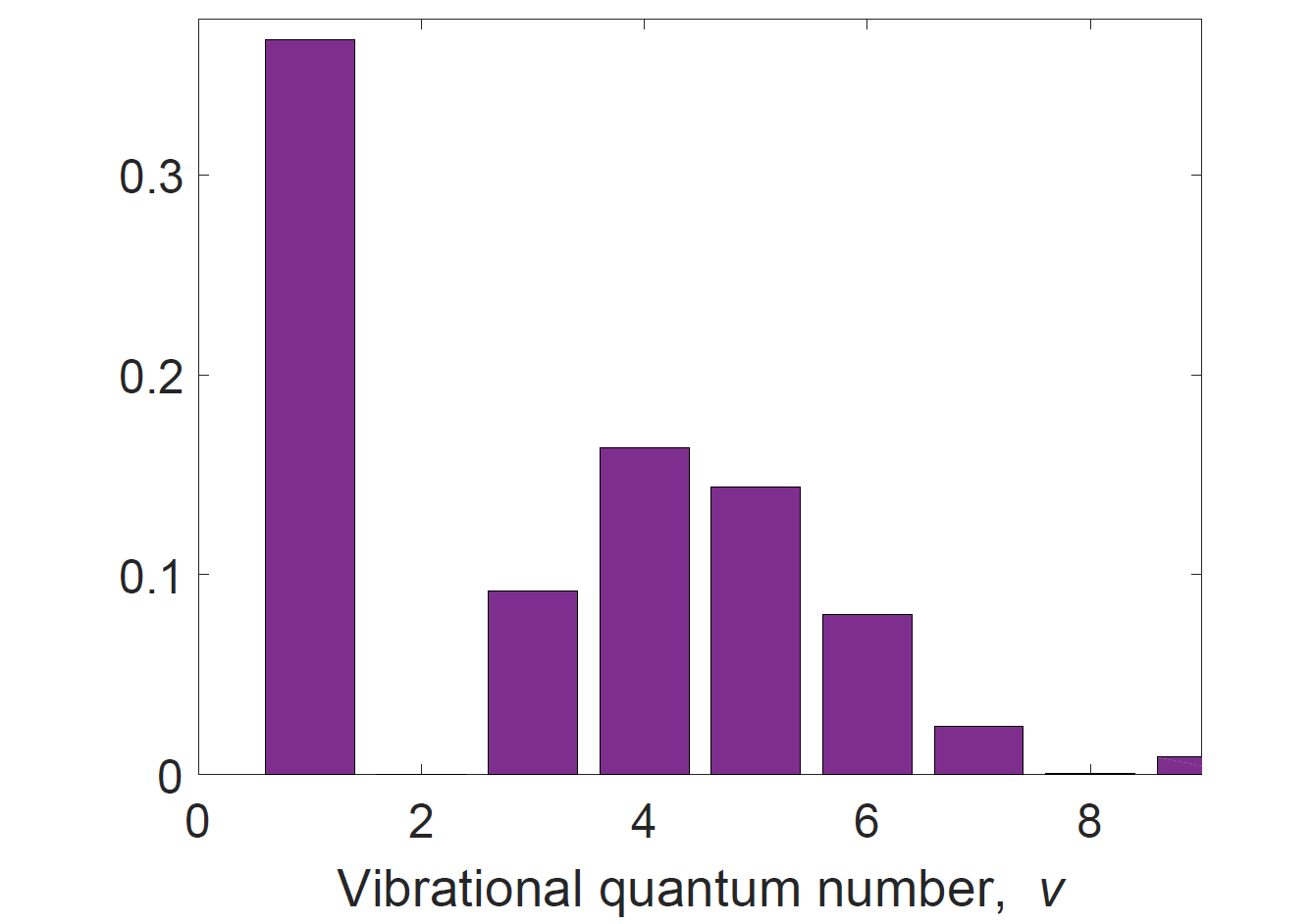}}
\subfigure[\ $\lambda = 0.75$]{
\includegraphics[scale=0.115]{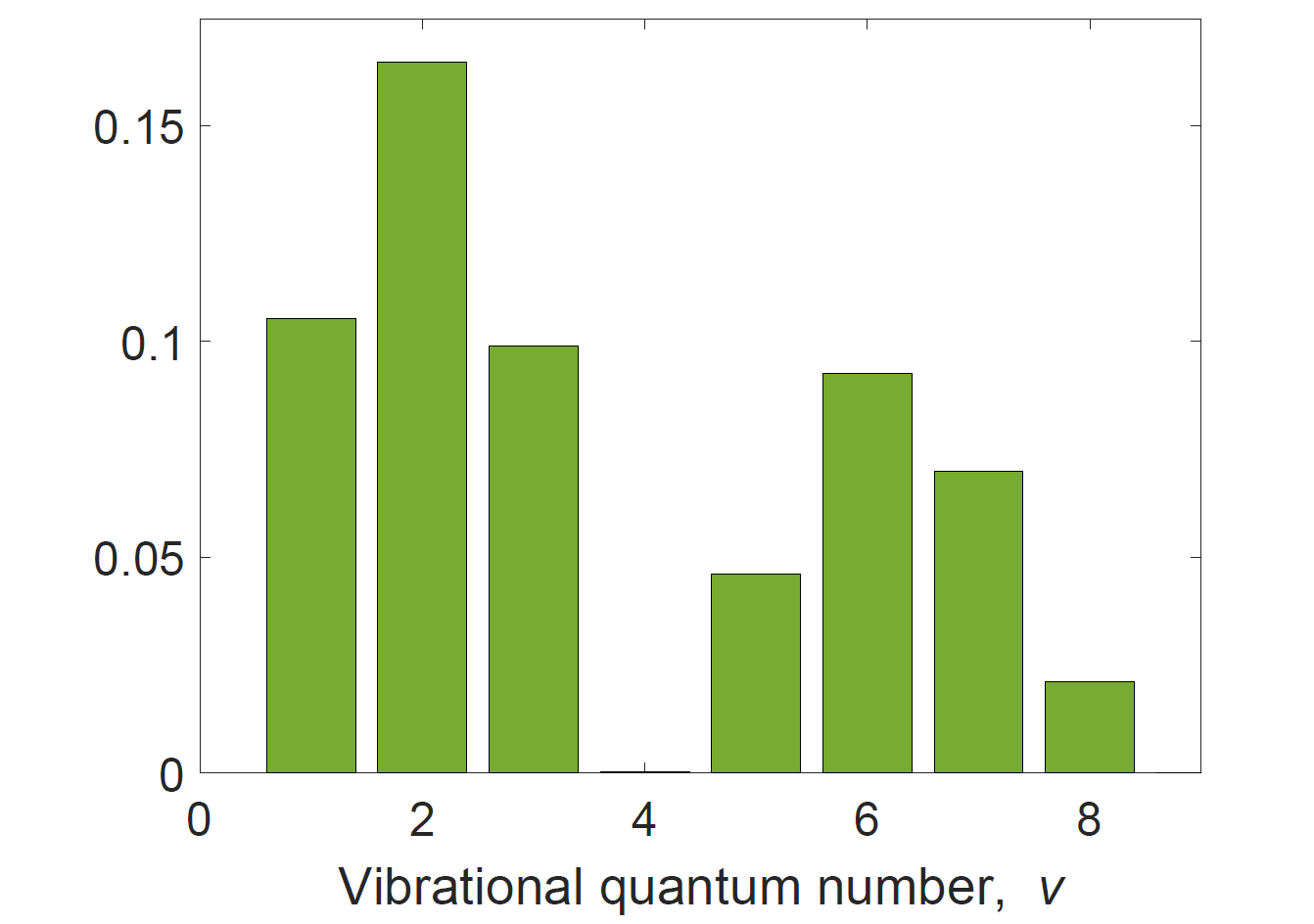}}
\subfigure[\ $\lambda = 1$]{
\includegraphics[scale=0.115]{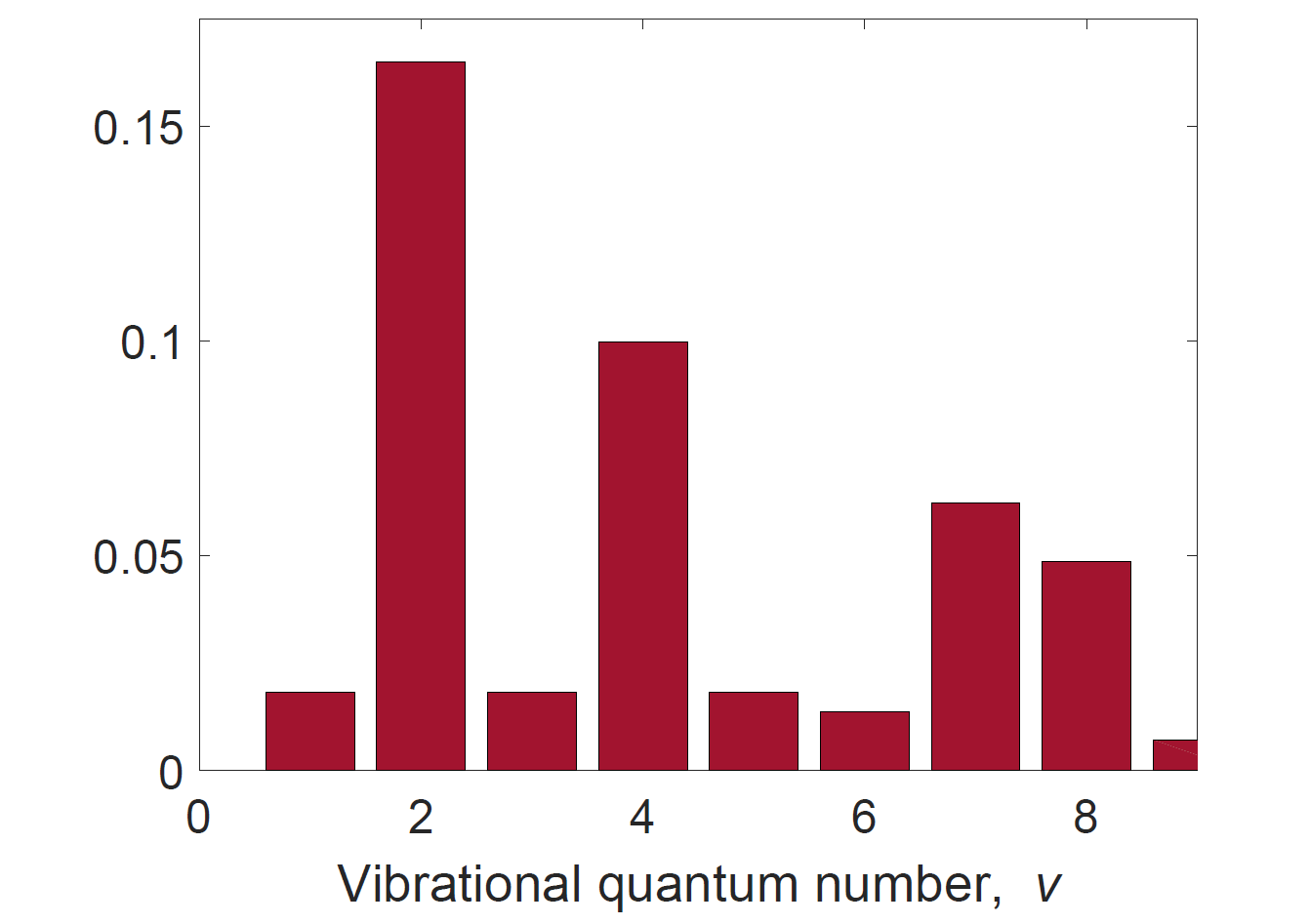}}
\subfigure[\ $\lambda = 1.25$]{
\includegraphics[scale=0.115]{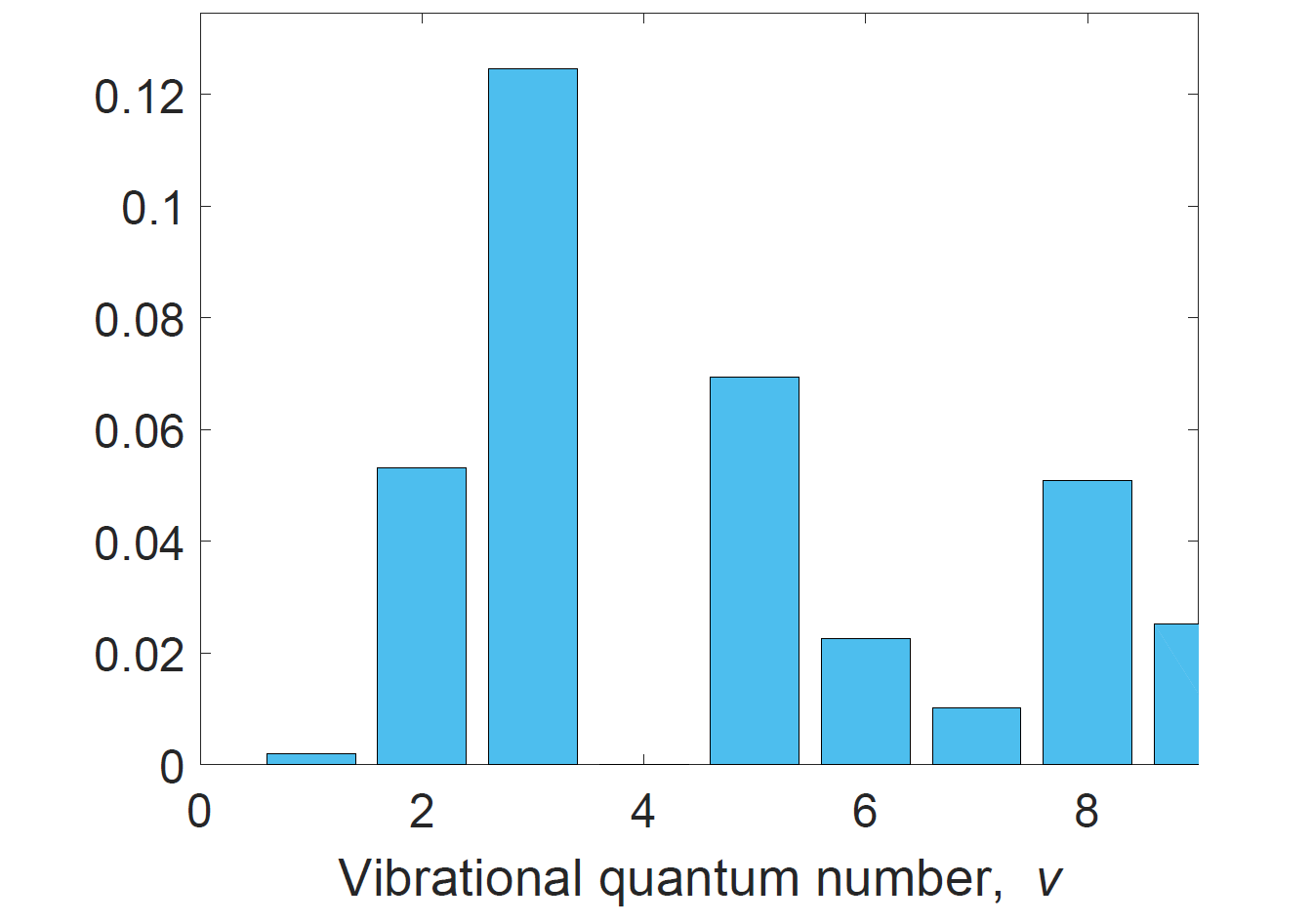}}
\caption{\label{figureAA} Square-moduli of diagonal matrix elements of $X_L^\dagger X_R$ operators for different values of $\lambda$ as used in Figure \ref{figure5a}.}
\end{figure}
Let us also consider the square-moduli of diagonal matrix elements of $X_L^\dagger X_R$ operators for different values of $\lambda$ (as used in calculation in Figure \ref{figure5a}) shown in Figure \ref{figureAA}. 
As discussed in Section \ref{sec:level31}, a two fold connection can be made between these values and the $I-V$ characteristics shown in Figure \ref{figure5a}. Firstly, the magnitudes can be roughly correlated with the overall magnitude of the calculated current, and secondly, the maximum conductance coincides with the bias voltage for which vibrational level $\lvert L, v\rangle$ corresponding to the maximum absolute value of $\langle v | X_R^\dagger X_L|v\rangle$ falls into the bias window. This further supports the interpretation presented in the paper.\\

\end{document}